**Historical Context, Scientific Context, and Translation of Haidinger's (1844)**

**Discovery of Naked-Eye Visibility of the Polarization of Light**

Robert P. O'Shea[1], Shelby E. Temple[2,3], Gary P. Misson[4,5], Nicholas J. Wade[6],

Michael Bach[7,8]

1. Institute of Psychology—Wilhelm Wundt, University of Leipzig, Leipzig 04109, Germany

2. School of Biological Sciences, University of Bristol, Bristol, BS8 1TQ, UK

3. Azul Optics Ltd, Bristol, BS9 4QG, UK

4. School of Life & Health Sciences, Aston University, Birmingham, B4 7ET, UK

5. Department of Ophthalmology, South Warwickshire NHS Foundation Trust, Lakin Road, Warwick, CV34 5BW, UK

6. Psychology, University of Dundee, Dundee DD1 4HN, UK

7. University Eye Center, Medical Center, University of Freiburg, Freiburg 79106, Germany

8. Faculty of Medicine, University of Freiburg, Freiburg 79106, Germany

Number of figures: 14

Number of tables: 0

Number of words: 4,895

**Conflict of interest**

There are no conflicts of interest.

**Author contributions**

ROS conceived the project and recruited NW, MB, GM, and ST. ROS made a first draft of the translation; MB, a native speaker of German, corrected it with advice from GM, ST, and ROS. NW and ST produced the figures for the translation. ROS wrote the first draft of the paper. All authors revised various drafts.

**Acknowledgements and Funding**

ROS is grateful to Urte Roeber for her support and advice about translation from German. We thank Sara Mae Stieb for helping to produce a flawless version of Haidinger's original German text. No specific funding was allocated to this translation.

**Corresponding author**

Please address all correspondence concerning this article to Robert P. O'Shea, Institut für Psychologie, Universität Leipzig, Neumarkt 9-19, Leipzig 04109, Germany. E-mail: robert.oshea@uni-leipzig.de

# Abstract

In 1844, the Austrian mineralogist Wilhelm von Haidinger reported he could see the polarization of light with the naked eye. It appears as a faint, blurry, transient, yellow hourglass shape superimposed on whatever one looks at. It is now commonly called *Haidinger's brushes*. To our surprise, even though the paper is well cited, we were unable to find a translation of it from its difficult, nineteenth-century German into English. We provide one, with annotations to set the paper into its scientific and historical context.

**Historical Context, Scientific Context, and Translation of Haidinger's (1844)**

**Discovery of Naked-Eye Visibility of the Polarization of Light**

**Preamble**

In 1844, the Austrian mineralogist Wilhelm von Haidinger ("Wilhelm Karl Ritter von Haidinger," 2020) reported he could see the polarization of light with the naked eye (Haidinger, 1844). It appears as a faint, blurry, transient, yellowish hourglass shape superimposed on whatever one looks at ("Haidinger's brush," 2020). It is now commonly called *Haidinger's brushes*.

Haidinger first noticed the phenomenon in various minerals that produce polarized light. He learned to see the phenomenon in the sky and in reflections from flat surfaces.

Haidinger's paper attracted the attention of some of the leading scientists of the time, including David Brewster (1851), George Stokes, James Clerk Maxwell (1850/1990), and Herman von Helmholtz (1910/1924). Helmholtz proposed that the human eye contains a polarizing structure, centered on the fovea and macula; this is now the consensus (Misson et al., 2020). The precise details of that structure remain a subject of research.

As far as we are aware, there is no translation into English of Haidinger's seminal paper. We provide one here with explanatory annotations setting the paper into its historical and scientific context.

**Annotated Translation**

II. *On the direct observation of polarized light and the position of the plane of polarization; by Wilhelm Haidinger.*

Soon after the discovery of the polarization of light by Malus, a mass of brilliant discoveries followed in this branch of human knowledge; the astuteness of Airy, Brewster, Biot, Arago, Fresnel, Herschel, Seebeck, and others has opened up so much of the beautiful and the remarkable in the highest splendor of colors that we can regard the richest harvests as having been reaped in this field. Every now and then a small harvest remains worth making, and as a contribution to such, a direct observation of polarized light with the naked eye, without the help of any instrument or any tool, should not be uninteresting. At the same time, the plane of polarization is unequivocally observed.

Upon careful observation of andalusite[1] plates cut parallel to the axis, I had long noticed that, on the whole, their color appeared to be very pale reddish, because they were cut so thin that not all of the red light was absorbed, but that one sometimes saw a flying phantom of a yellowish color,[2] which disappeared again when the plate was looked at more carefully. Cut  perpendicularly to the optical axes, when one looks in the direction of these axes, the andalusite shows beautiful bundle of pale green light rays {Lichtbüschel}[3] in directions depending

---

[1] Andalusite is common, metamorphic, aluminium nesosilicate mineral. Clear forms, of the sort Haidinger used, are pleochroic: reds, greens, and yellows can be visible from different angles. Andalusite. (2020). In. Wikipedia. https://en.wikipedia.org/w/index.php?title=Andalusite&oldid=975229499

[2] This "flying phantom of a yellowish color" is what came to be known as Haidinger's brushes: two faint, blurry, yellow, mirror-symmetrical shapes, together rather like an elongated symbol for infinity, centred on where we look. Each brush is about 2° of visual angle in length. They may be accompanied by two, similar, blue brushes, with an axis perpendicular to that of the yellow ones.

[3] We give the original German in {braces} when our translation might be a little too free or controversial. By itself, the "Büschel" part of "Lichtbüschel" can mean "tuft" or "tufts" or "bundle" or "bundles", or "sheaf" or "sheaves" (as of cut stalks of wheat tied tightly in the middle). One of the early English scientists who researched the phenomenon, possibly George Stokes (https://en.wikipedia.org/wiki/Sir_George_Stokes,_1st_Baronet), translated "Büschel" as "brush(es)", and the rest, as they say, is history. In his paper, Haidinger used the three meanings interchangeably.

By the way, "Büschel" cannot now be translated as "bushel" (e.g., of wheat); it is a false friend. Nowadays, "bushel" means a measure of dry grain of about 36 litres, although numerous images similar to some of Haidinger's figures of the brushes appear if one searches the internet for on the crystal structure, enclosed by two dark red spaces. Neither looking directly, nor with a magnifying glass, was I able to find a trace of the yellow-colored figure.[4]

[p. 30][5] On a later occasion, I tried to see a difference in intensity between the two images, produced by Iceland spar [transparent calcite],[6] of a black square on a white background, and because this is almost imperceptible, it was necessary to view one and the other alternately for comparison.[7] Soon yellowish and

---

"bushel of wheat". And even though these images, of a sheaf of cut stalks of wheat, have approximately the correct color, they are depicting a three-dimensional object, which the brushes are not.

For simplicity we have translated "Büschel" as "tufts" because Haidinger may have been thinking of a "Gamsbart", a flat tuft of hair worn on alpine hats in Austria and in Bavaria (https://en.wikipedia.org/wiki/Gamsbart).

In these footnotes, we refer to them as "brushes", because that is how they have become known. Finally, in this case, we have translated the compound word "Lichtbüschel" as "bundle of light rays" because this expression was common in English-language papers about optics in the early nineteenth century.

[4] This is because Haidinger's brushes are entoptic phenomena, that is, produced in the eye. Like other entoptic phenomena, for example, the shadows of the blood vessels lying across the surface of the retina, the brushes fade to invisibility within a few seconds if the stimulus producing it is unchanging. Because Haidinger was looking intently for the brushes, he was presumably not changing the orientation of his eyes relative to the orientation of polarization of light transmitted through the crystal, yielding the fading.

The strength of evidence supports the hypothesis that Haidinger's brushes arise from the effects on light from passing through macular pigments (lutein, zeaxanthin, and meso-zeaxanthin) in the Henle fibres—the axons of the photoreceptors that radiate outwards from the centre of the fovea like the spokes of a wheel. Macula pigments and the nanostructure of the Henle fibres combine to give preferential absorbance of polarized light with its electric vector perpendicular to the axes of fibres. The density of macular pigments is greatest centrally and decreases peripherally to low levels at about 4° of eccentricity.

Any light coming into the eye, in this case from the andalusite, has to pass through the macular pigments and the Henle fibres before reaching the outer segments of the photoreceptors, containing the photopigments that transduce light into electrical signals. Incoming, linearly polarized light in one orientation, say vertical, will be maximally transmitted to those photosensitive pigments under vertically oriented fibres (i.e., containing horizontally oriented macula pigment molecules), and minimally transmitted to those under horizontally oriented fibres (i.e., containing vertically oriented macula pigment molecules).

The yellow color of the brushes is because macular pigments strongly absorb violet-blue-turquoise light (400–520 nm) leaving the rest of the visible spectrum green-red (520-730 nm) which combined appears yellow.

[5] These sort of insertions (in this case "[p. 30]") give the place in the text in which a new page begins in the original paper (in this case, p. 30).

[6] Words we give in [square brackets] are our insertions for clarity. In this case, Iceland spa is a transparent calcite crystal that produces two images of an object, each one with an orthogonal plane of polarization.

[7] Here, Haidinger discovered that moving his eyes, in this case to light that is orthogonally polarized, restored and enhanced visibility of the brushes. A similar technique of moving the eyes,

grayish-violet colors emerged, which became more and more clear as complementary colors,[8] until they finally took on the shape of yellow light tufts {Lichtbündel} on a violet-gray background. As shown in Fig. 1,[9] the upper ordinary [10] image O [appeared as] horizontal tufts {Büschel}, the lower extraordinary image E[11] as vertical tufts, narrow in the middle, diverging on both sides.

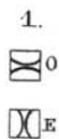

[Figure 1.   **Appearance of the yellow tufts after alternately looking at the two images of a black square on a white background produced by Iceland spar. O shows the ordinary image, E shows the extraordinary image.]**

But black means no light {Schwarz ist aber der Abgang des Lichtes}; two white pictures of an opening cut out in black paper, viewed through Iceland spar, or the two pictures [viewed] through a dichroscopic magnifier [dichroscope],[12] held up to the brightest light of the same kind, easily gave the appearance of Fig. 2. The yellow tufts of the ordinary ray *O* [were] vertical, those of the

---

although in this case with unpolarized light, had been reported by Charles Wheatstone in 1830 for seeing the entoptic shadows of the retinal blood vessels.

[8] Here, Haidinger recognized that while looking at one image, for example, showing faint, vertical, yellow brushes, an afterimage of that color, blue, developed. When he looked at the other image's faint horizontal brushes, the afterimage color then summed with the background, making it appear blue, augmenting the visibility of the yellow, horizontal brushes through simultaneous color contrast. At the same time, a second, horizontal, blue afterimage developed. This then augmented the visibility of the yellow, vertical brushes when he looked again at the first image, and so on.

[9] All of Haidinger's figures were given on a plate (Plate 2) at the end of the volume of *Annalen der Physik*. Here we give them separately, with figure captions we wrote, as is now the practice.

[10] In Haidinger's time, an ordinary image was the one that followed Snell's law (https://en.wikipedia.org/wiki/Snell%27s_law), specifying the relationship between the angle of incidence of a ray of light falling onto a transparent surface, say of glass, and the angle of refraction. For example, ordinary images that come from rays of light that are perpendicular to a crystal's surface are not refracted at all.

[11] The extraordinary image was the one that broke Snell's law. It is extraordinary because it meant that a ray of light that is perpendicular to the surface of a crystal was refracted.

[12] A device Haidinger invented to allow one to see magnified versions of the two images from any polarizing material (https://de.wikipedia.org/wiki/Dichroskop). The magnification made it easy for Haidinger to move his eyes from one oppositely polarized image to the other; without magnification the two images are very close. Of course, the magnification would have no optical effect on the size of the brushes, which arise from a structure on the retina.

extraordinary ray *E* [were] horizontal, when the axis of the rhombohedron was in a vertical plane. But now the ordinary ray is polarized in the plane of the main rhombohedron section, the extraordinary ray is polarized perpendicular to it. *So, the orientation of the light tufts shows exactly the orientation of the plane of polarization.*[13]

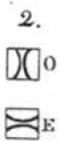

[Figure 2.  **Appearance of the yellow tufts after alternately looking at the two images of a white square on a black background produced by Iceland spar. O shows the ordinary image, E shows the extraordinary image.**]

The examination of the andalusite plates now had to be resumed, but no longer at the distance of clearest vision, but held straight in front of the eye to see through against bright light. Now the tufts showed themselves in every direction in which one looked, but always regularly parallel to the vertical axis. From this position of the tufts it follows, in comparison with Fig. 2, *that the remaining light green ray in andalusite is the ordinary* [p. 31] *one*. In fact, in the dichroscope, the upper image *O* remains light green, the lower one *E* becomes dark red. The vertical position of the tufts can be observed on the surfaces ∞ *A*, and ∞ ˘*D*, and ∞ *D̄* of a prism of 90° 51′ and its two diagonals. Indeed, one can perceive them {man nimmt sie … wahr} on unpolished or polished glacial pebbles [of andalusite] held close to the eye. The observation is made very clearly when the crystal is first held vertically and gradually rotated to the horizontal, because the retina has become more sensitive due to the complementary color. This rotation of the polarizing plane around the line of sight as an axis is not always possible.

---

[13] In Haidinger's time, orientation of the plane of polarization was governed by the axes of the crystal that produced the polarization. Nowadays, it is given by the orientation of the electrical field of light, and is orthogonal to the axis of a crystal.

With stationary polarized light one inclines the head alternately to the right and left, and then immediately perceives the tufts clearly, even if in a somewhat different orientation.[14]

Next, of course, was a tourmaline[15] plate, Fig. 5[16]. It showed very clear tufts of light in a horizontal position against the vertical axis *AB*. *The light beam passing through, compared with Fig. 2, is therefore the extraordinary one from the position of the tufts*. Indeed, even in the dichroscope, the ordinary image *O* is black, everything *O* is absorbed, and the extraordinary image *E* appears with a light brown color.

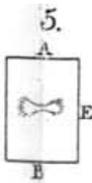

[Figure 5. **Appearance of yellow tufts in a plate of tourmaline. Because the tufts are horizontal to the vertical axis of the plate (AB), it must be from the extraordinary image (E).**]

If one compares the andalusite, Fig. 3, which allows the ordinary ray to pass, and the tourmaline, Fig. 5, which absorbs the ordinary ray, it naturally follows that they combine to transmit only the *minimum* of light corresponding to the thickness of both. Crossed, they let the *maximum* through. In andalusite the ordinary ray is the brighter, stronger one, in tourmaline the extraordinary ray [is the brighter, stronger one].

---

[14] This small change in orientation may be from corneal birefringence, which has since been found to occur in the cornea due to stromal layers that cause a specific and unique orientation of fast and slow axes for each individual.

[15] https://en.wikipedia.org/wiki/Tourmaline

[16] We have retained Haidinger's original figure numbers, in this case 5, which are only approximately ordinal against place in the paper. For example, Figure 3 appears next. Figure 4, which nowadays would come after Figure 3 appears in Plate 2, but was for another of Haidinger's papers in the same volume of *Annalen der Physik*. The same goes for Figure 6, and so on.

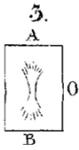

**[Figure 3.   Appearance of yellow tufts from the ordinary image (O) in a plate of andalusite.**

The green tourmalines also clearly show the yellow tufts; only the almost colorless, pale rose-red ones from Elba allow both rays to pass through. Through the dichroscope, [p. 32] only the ordinary image appears a little more colored.

It was now interesting to examine the polarized light obtained by the usual methods; everywhere one could more and less clearly notice the yellow tufts with the accompanying bluish spaces.

A black mirror held horizontally under the eye, Fig. 7, [such as] the broad surface of a well-polished table of the same color, even a shiny waxed floor, shows vertical tufts, namely in the main polarizing section[17]. In vertical window casements one easily observes almost horizontal tufts, which are only slightly inclined in the plane of polarization, when one alternately views the reflected image of an initially adjacent horizontal window bar or cross, which also renders the retina more sensitive to the percept.

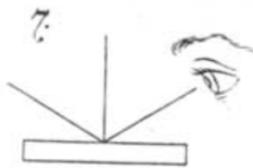

**[Figure 7.   Plan view of how to see the tufts in polarized light reflected from a dark, polished surface.]**

---

[17] This must refer to the angle of incidence of unpolarized light on a surface that produces a reflection that is perfectly polarized. This angle is known as Brewster's angle (https://en.wikipedia.org/wiki/Brewster%27s_angle) after David Brewster, who published it in 1815. Haidinger lived for two years in Edinburgh, from 1823. He must have had considerable contact with Brewster, assisting with Brewster's *Edinburgh Journal of Science*.

Light that has passed obliquely a layer of glass plates, even through a single glass plate, shows the yellow tufts, the former naturally more clearly in the direction *perpendicular to the plane of incidence* of the rays.

The blue sky is clearly polarized in principal planes, which are great circles through the sun. As far as the optics are concerned, the atmosphere is a kind of uniaxial {einaxigem} crystal, which in the direction of the axis has the color of the sunlight from which it is illuminated, in the direction perpendicular to it the color of the infinite depth of space, omitting all light, black, modified by the physicality of the illuminated atmosphere to more and less dark blue.[18] If you quickly look somewhere at the blue sky, there appear clearly, almost like two delicate yellow, interconnected spots with an apparent size of about 2° [each], the yellow tufts in the direction of the [p. 33] principal plane [of the sky crystal]. The impression is slightly intensified by closing the eyes quickly and opening them at the same height on a great circle through the sun, perpendicular to the first one. On the retina, the second tufts appear rotated by 90° against the first, and much more clearly. The change puts the wonderful phenomenon in its full clarity, without any apparatus other than the naked eye.

The stronger the light is polarized, the more reliably the tufts appear, most clearly in crystals which absorb one ray, and usually appear colored. Even in completely transparent Iceland spar, the ordinary ray is slightly more absorbed than the extraordinary one. In rather thick pieces seen obliquely through two parallel surfaces, so that the ray is directed rather perpendicular to the axis of the rhombohedron, one perceives {nimmt} the yellow tufts in the position *perpendicular* to the axis, namely the more intense one {Ueberschuß}, the stronger beam, the extraordinary one. The yellow tufts appear much more clearly in a honey-yellow Iceland spar from St. Denis in France. But this Iceland spar is reddish-honey yellow in the direction of the axis, even to the naked eye; perpendicular to the axis it is more yellowish. Separated by the dichroscope by looking vertically at the axis is reddish honey yellow, pale wine yellow, and much brighter than *O*.

---

[18] This implies that Haidinger had an appreciation of the necessity for scattering of light to give the sky its blue colour. This was a very old idea, explained by Ibn Al-Haytham in the early eleventh century Alhazen. (ca. 1024/1989). *The optics of Ibn Al-Haytham: Books I–III: On direct vision* (A. I. Sabra, Trans.). The Warburg Institute. (Original work published ca. 1024). A translation of some of  Alhazen's writings from Arabic into Latin was published in the sixteenth century, so would have been available to Haidinger, although it did not include the material on scattering. The theory of scattering was published by Lord Rayleigh (John William Strutt) in 1871.

The light-colored variety of cordierite[19] shows three colors perpendicular to each other in ordinary light, namely in Fig. 8, which represents a cut cube in the Royal Mineral Collection [of Vienna]. [One side] is beautifully blue (P), [another] is yellowish clear gray (M), [and the third is] a somewhat cloudy, slightly bluish gray (T).

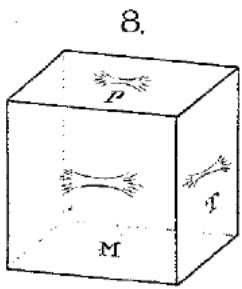

[Figure 8.   Illustration of a cube of clear cordierite showing how the tufts appear on its three orthogonal surfaces.]

Through the dichroscope the colors are decomposed:

    *P* into dark blue and very pale blue,

    [p. 34] *M* into dark blue and yellowish gray,

---

[19] https://en.wikipedia.org/wiki/Cordierite

*T* into yellowish grey and very pale blue.

The tufts lie as shown in the figure. They show the brightness of the colors. The lightest tint is the yellowish-grey, the weakest, darkest, is the dark blue, because it is first absorbed. The blue color of P does not contain yellow in the mixture; the tufts appear distinct but slightly purple.

In the case of barite {Schwerspath},[20] this most peculiar species that deserves its own monograph in terms of color alone, the tufts are not the same in all varieties. The yellow ones from Felsöbánya[21] show them [the tufts] like Fig. 9 in the large diagonal of the rhombus. — But the upper ordinary image O is also pale straw yellow, lighter than the lower lemon-yellow image E. In a pale clove-brown one from the Imperial-Royal Court Mineral Collection, from Beira in Portugal, the tufts lie in the small diagonal of the rhombus. But the upper image O is also dark violet-blue {violblau}, the lower image E is pale wine yellow, therefore lighter, and the effect of the light corresponding to this color remains, while the dark violet-blue is absorbed.

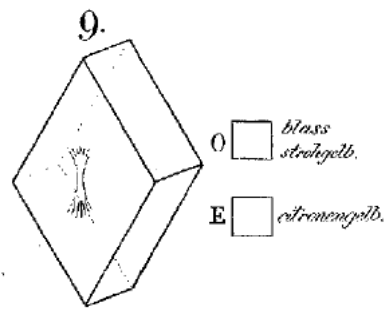

[Figure 9.   Illustration of a barite crystal with tufts on the main diagonal. The ordinary O image has a pale straw yellow color, the extraordinary image E has a lemon-yellow color.]

---

[20] https://en.wikipedia.org/wiki/Baryte
[21] Now Baia Sprie in present-day Romania.

Mica flakes,[22] thick enough so one ray turns out to be much brighter than the other, which therefore becomes more strongly absorbed, show the tufts in ordinary light in the direction of a plane perpendicular to the plane of the axes through the center line. The thinner the flakes, the smaller the absolute difference between the effect of the two perpendicularly polarized rays. In ordinary light, no more tufts are seen. They appear more beautifully and vividly when you look through thin mica flakes or even through birefringent flakes [when] looking at a polarized light surface, but with a strange modification.[23]

[p. 35] If polarized light from a horizontal surface is viewed through a device polarizing in a certain direction, the orientation of the tufts {Lichtbüschels} after the latter is not changed. If [the orientations of polarization] of both are parallel, for example if both are vertical, the maximum of light enters the eye; if the [orientation of polarization] of the device is horizontal, the minimum of light enters the eye, because the vertically polarized beam has been absorbed by the surface which reflects or transmits only a horizontally polarized beam. The maximum absorption occurs when the analyzing apparatus is rotated in azimuth by 90°. While rotating once by 360°, at 90° there is a maximum, at 180° a minimum, at 270° a maximum, at 360° or 0° again the initial minimum of absorption.

---

[22] Mica acts as a waveplate or a retarder, altering the polarization of light travelling through it. They are commonly used in mineralogy to determine refractive indices. (https://en.wikipedia.org/wiki/Waveplate)

[23] We think that from here, Haidinger was describing the effect of rotating a mica flake (plate) between the eye and a source of linear polarized light. The mica flake is acting as a retarder altering the appearance and orientation of the brushes. We do not know how thick the mica was (and hence its retardation) but it seemed to be acting as a full-wave retarder (i.e., close to the wavelength of light).

The situation is different if a flake is used that transmits two equal or similarly intense rays polarized perpendicularly to each other, as is the case with crystallised flakes everywhere, except in the axial directions.

We have to distinguish three different positions of the yellow tufts, 1) the vertical one on line $AA_1$, Fig. 11, which, like the polarizing horizontal plane itself, remains fixed;

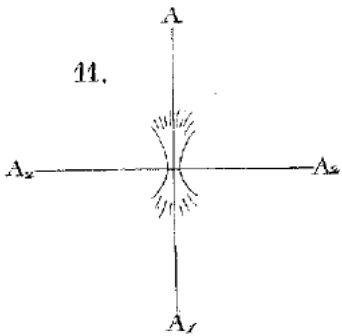

[Figure 11. Appearance of yellow tufts from horizontally polarized light viewed through a mica flake rotated to a particular angle.]

2) the two polarized tufts $BB_1$ and $B_2B_2$, Fig. 12, which remain in directions perpendicular to each other for each of the two polarizations when the opposite one is cancelled by an adjusting apparatus;

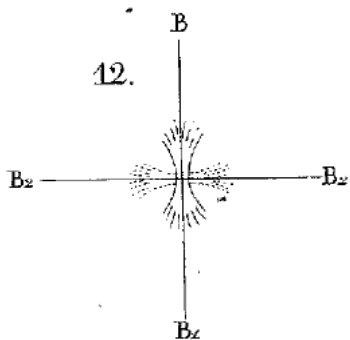

[Figure 12. Appearance of two sets of yellow tufts from horizontally polarized light viewed through a mica flake rotated to different angles. Flake $BB_1$ is vertical; flake $B_2B_2$ is horizontal.]

3) the tufts $CC_1$, Fig. 13, which are observed when $A$ is covered by $B$ in different azimuths [rotations].

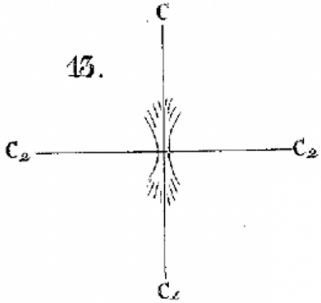

[Figure 13.  Appearance of the yellow tufts when viewed through mica flakes A and B at 90° angles.]

When flake {Platte} $BB_1$ is placed on $AA_1$, the tufts also appear vertically.

If one rotates $BB_1$, Fig. 14, at the top against [p. 36] the right side [clockwise] by an angle $\varphi$, the yellow tufts appear in the direction $CC_1$ with the line $AA_1$ enclosing an angle $2\varphi$.

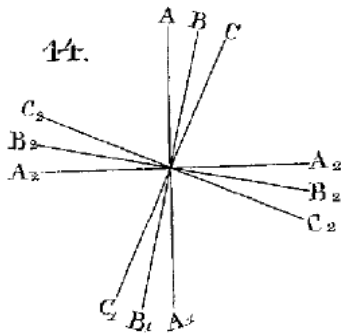

[Figure 14.  Orientations of the tufts at three angles of rotation of the same mica flake.]

The angular distance of the real plane of polarization from the fixed one is twice as large as the distance of the original plane of polarization of the one beam in the flake. It advances by the angle of rotation $\varphi$.

When one rotates the flake by 45°, $CC_1$ becomes horizontal. The deviation of the plane of polarization is then maximal, and therefore gives, when examined by a firmly polarized apparatus, the absorption phenomena opposite to position $AA_1$, although the flake has been turned only 45°.

When turning $BB_1$ by 90°, the yellow tufts assumes position $A_1A$, that is, it is vertical, but they are turned upside-down. The phenomena of absorption are as in the original [prior] position.

With a rotation of 180°, the appearance of the yellow tuft falls completely back in the direction $AA_1$. The tufts have completed a rotation of 360°, although the flake only described a semicircle. When the latter has been rotated through 360°, the visible tufts have already rotated twice.[24]

With the tufts, the alternating lightness and darkness of the crystal plates, distributed according to octants[25] in the polarized and then analyzed light, are directly connected; the effect opposite to perfect polarization increases on either side of the line determined by the position of the tufts, and is maximal perpendicular to that line. But there is the second plane, through which the polarization perpendicular to the first [p. 37] takes place. If one or the other line covers the tufts of the original polarization of the mirror, the tufts that appear must likewise become vertical. The actual maximum of the opposite effect falls between the two positions and therefore returns four times, namely at 45°, 135°, 225° and 315°. If the polarizing and the analyzing light bundle are positioned in parallel, these spaces are bright, the 0°, 90°, 180°, 270° in between are dark, and if the bundles are crossed, the latter are light, the former dark.

Direct observation on the plane of polarized light can easily be determined by the yellow tufts one discovers [when looking at the light] through mica flakes.

---

[24] This is a property of waveplates (https://en.wikipedia.org/wiki/Waveplate).
[25] We think this refers the property of rotation of a waveplate relative to a polarizer: it will cause eight changes in light and dark every 45° of rotation.

After a brief glance through the mica and then quick removal, the retina remains more sensitive to the percept of tufts in a different direction. But this is not a mere complementary percept, for the tufts show themselves on the polarized background in a firmly determined direction, whatever the situation may have been seen through the mica.

The appearances described above of tufts at twice the angular distance for rotations of the mica flakes can also be compared differently. Define the position of the mica flake, including $BB_1$, Fig. 15. If you now move the original polarization plane $AA_1$ away from $B$ towards the left, the tufts appear at the same angle to the right in the direction $CC_1$. It is a true circular, or rather gyroidal[26] polarization. The angle $B\,M\,C$ is always equal to the angle $B\,M\,A$; $C$ is to the right of $B$ when $A$ is to the left; conversely, if $A$ is rotated to the right, $C$ is to the left of $B$.

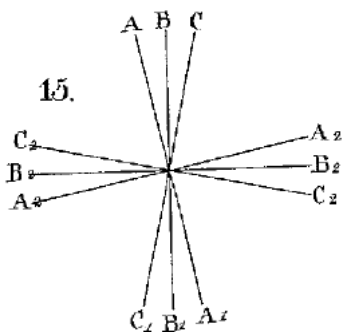

[Figure 15.  Orientations of the tufts at three angles of rotation of the same mica flake.]

In the middle of the yellow light tufts, which are detected by mica flakes on the surfaces of polarized light [placed in the optical path of linearly polarized light], direct observation on these can be easily obtained.

Similar observations cannot be made with [p. 38] the usual polarization instruments with a fixed polarization mirror. They are very easily carried out with those which are set up according to the principle of Fig. 16. The rays that differ due to polarization are directed again in the direction of the axis of the tubes by parallel metal mirrors.

---

[26] A term used in mineralogy to describe light that has been polarized spirally to the right or left.

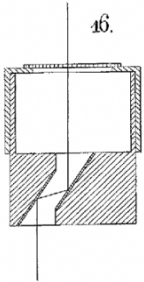

**[Figure 16. Side view of apparatus holding a crystal at the top and front-surfaced mirrors at the bottom through which the effects on a beam of light passing through the crystal can be observed.]**

Only in the direction of the axes do the tufts reappear in a fixed position.

But what, finally, are the yellow tufts which were observed in every [orientation of] polarized light? If light can be explained as wave-like movement of ether particles, considered by Dr. Young[27] in analogy to the phenomena of a vibrating rope according to Herschel [(1831)], then the impressions of the waves must be observed on both sides of the cross-section in the plane of vibration at the greatest deviation, Fig. 17, because they stand still for a moment, only to return, as is so clearly seen in {wahrnimmt} vibrating strings.

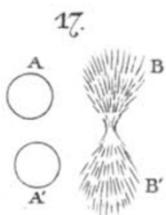

**[Figure 17. Illustration of how the tufts appear (BB') from vibration of an etheric wave in plane AA'.]**

---

[27] In 1800, Young used a magnifying glass to look at vibrating piano strings surrounded by reflecting foil.

The yellow tufts have a shape according to this analogy. *The eye sees in the polarized light the vibrations of the etheric particles*, if such a statement is not too daring. Then the plane of oscillation of the waves is clearly established by the phenomenon. According to this, however, in a beam polarized by reflection, the vibrations are not, as is assumed as the basis of calculation [(Müller, 1843, vol. 2 p. 274)], parallel to the plane of the polarization mirror, but are perpendicular to it. The plane of oscillation of a beam which has been polarized by a [p. 39] tourmaline plate is not parallel to the crystallographic main axis of the tourmaline plate, but is perpendicular to it.

Even if I could wish to investigate this apparent contradiction in more detail here, and to pursue the influence of the observation of the tufts in polarized light, especially with regard to circular polarization, for the time being is too far-reaching, so I would invite the scientists of optics {Optiker} to visit this remarkable phenomenon and include it in their studies.[28]

### [Haidinger's references]

[J. F. W. Herschel] Of light. [Translated by {D.} J. C. E. Schmidt, 1831, p. 538. [Cotta'sche Buchhandlung, Stuttgart / Tübingen Harvard.]

[Müller, J. (1843).] Pouillet's Physics, by Müller, vol. 2 p. 274. [Johann Heinrich Jacob Müller. (1843). Pouillet's Lehrbuch der Physik und Meteorologie, für deutsche Verhältnisse frei bearbeitet [Pouillet's textbook of physics and meteorology, freely edited for German conditions]. Vol 2. Friedrich Vieweg und Sohn, Braunschweig.]

---

[28] Indeed they did. Scientists who researched Haidinger's discovery included David Brewster, George Stokes, James Clerk Maxwell, and Herman von Helmholtz.